\documentclass{aa}  
\usepackage{graphicx}
\usepackage{txfonts}
\usepackage{amsmath}
\usepackage{amssymb}
\usepackage{listings}
\usepackage{xcolor}
\usepackage[colorlinks=true,linkcolor=blue,citecolor=blue, filecolor=blue, urlcolor=blue]{hyperref}
\begin{document}
\title{Understanding observational characteristics of solar flare current sheets}

\author{Zining Ren \inst{1,2}
        \and Yulei Wang\inst{1,2}
        \and Xin Cheng\inst{1,2}
        \and Mingde Ding\inst{1,2}
       }

\institute{School of Astronomy and Space Science, Nanjing University, 
           Nanjing 210023, People's Republic of China \\
           \email{wyulei@nju.edu.cn,xincheng@nju.edu.cn}
           \and
           Key Laboratory for Modern Astronomy and Astrophysics (Nanjing University), 
           Ministry of Education, Nanjing 210023, People's Republic of China
          }
 
\abstract
% context heading (optional)
{The elongated bright structures above solar flare loops are suggested to be current sheets, where magnetic reconnection takes place.
Observations have revealed various characteristics of the current sheet; however, their physical origin remains to be ascertained.
}
% aims heading (mandatory)
{In this study we aim to reveal the relations of observational characteristics of current sheets with the fundamental processes of magnetic reconnection.
}
% methods heading (mandatory)
{Using high-resolution 3D magnetohydrodynamic simulations of turbulent magnetic reconnection within a solar flare current sheet, we synthesized the remote-sensing observations of the current sheet and determined their physical properties.
}
% results heading (mandatory)
{Turbulent magnetic reconnection can significantly broaden the apparent width of the current sheet, which is much larger than the realistic physical width because of the superposition effect. 
The differential emission measures of the current sheet have two peaks; the high-temperature component is spatially related to confirmed small-scale reconnection sites, showing that the current sheet is directly heated by reconnection. 
Moreover, we demonstrate that strong turbulence can cause the nonthermal broadening of spectral lines at both the current sheet and flare loop-top regions. A strong correlation between them in time is also observed.}
% conclusions heading (optional), leave it empty if necessary 
{Our 3D turbulent magnetic reconnection flare model can be used to interpret primary observational characteristics of the elongated bright current sheets of solar flares. 
}


\titlerunning{Observational characteristics of solar flare current sheets}
\authorrunning{Ren, Zining, et al.}
\maketitle
%-------------------------------------------------------------------
\section{Introduction}
Magnetic reconnection is a fundamental physical process occurring throughout the Universe, for example in black hole jets \citep[e.g.,][]{Yang2024}, the solar corona \citep[e.g.,][]{Yan2022,cheng2023nc}, and the Earth's magnetosphere \citep{Huang2024}.
During magnetic reconnection, two sets of magnetic field lines with opposite polarities are brought together, forming a dissipation region characterized by a strong current in the center, known as the current sheet.
Theoretical studies have demonstrated that the physical properties of current sheets determine the development of magnetic reconnection \citep{2000mare.book.....P}. 
Therefore, experimental and observational studies on current sheets are crucial for understanding magnetic reconnection processes and thus the energy release mechanisms of eruptive phenomena.

In laboratories, current sheets can be generated via interactions between strong lasers and solid targets \citep{2023SSRv..219...76J}.
\citet{2023NatPh..19..263P} recently reproduced the features of turbulent reconnection in a laser-driven current sheet.
Through in situ observations, kinetic-scaled processes and structures of magnetic reconnection taking place in the Earth's magnetosphere have been widely investigated.
For instance, with in situ measurements from the Magnetospheric Multiscale (MMS) mission, \citet{2023NatAs...7...18W} show that bursty reconnection in the solar wind is more common than previously expected, and that it actively contributes to solar wind acceleration and heating. 
Unlike experimental and in situ measurements, solar observations can provide remote-sensing data of solar current sheets, from which their formation and macroscale properties can be studied.
Solar current sheets appear as an elongated bright structure above the loop top of flares as often observed by high-temperature passbands of the Solar Dynamics Observatory Atmospheric Imaging Assembly (SDO/AIA) 131$\AA$ (11\,MK) and AIA 94$\AA$ \citep[7\,MK;][]{2011ApJ...727L..52R,2011ApJ...732L..25C}. 
Via the differential emission measure (DEM) technique, it was further determined that the temperature of current sheets is much higher than that of the background, sometimes even reaching $20\,\mathrm{MK}$ during strong flares \citep[][]{2018ApJ...866...64C}. 

The thickness of current sheets is a crucial parameter for determining the efficiency of magnetic reconnection \citep{2000mare.book.....P}.
Theoretical analysis and kinetic simulations indicate that the thickness is on the order of the proton gyro-radius \citep{1996ApJ...462..997L,Drake1997}, translating to tens of meters in the solar coronal environment.
In solar observations, the thickness of current sheets is generally inferred from the brightness distribution.
The full width at half maximum (FWHM) of the brightness distribution perpendicular to the current sheets suggests a thickness of about $10\,\mathrm{Mm}$--$25\,\mathrm{Mm}$ \citep[see][]{2018ApJ...853L..15L,2018ApJ...866...64C}, which presents a considerable discrepancy with theoretical predictions.
Some numerical studies show that, because of the thermal conduction, the heat generated in the core reconnection region might leak into the adjacent region, causing a halo that appears to broaden the current sheet \citep{2009ApJ...701..348S,yokoyama2001}.
For current sheets with a large aspect ratio, the development of tearing-mode instability (TMI), which produces various magnetic flux ropes, was also proposed to play a role in the broadening \citep{Mei2012}.
Turbulent magnetic reconnection, proposed first by \citet{1999ApJ...517..700L}, is especially believed to be able to significantly increase the current sheet thickness; this has been validated by numerical simulations performed under relatively ideal plasma and boundary conditions \citep[see][]{Kowal2009,Kowal2017,2020ApJ...901L..22Y,Daldorff2022,2022ApJ...940...94B}.
However, in the realistic coronal environment, how turbulent magnetic reconnection affects the thickness of current sheets remains unclear. 

Moreover, small-scale structures were also detected to be ejected out of the current sheet. 
Using Large Angle and Spectrometric Coronagraph (LASCO)/C2 images, \citet{2013ApJ...771L..14G} identified plasmoids in a post--coronal mass ejection current sheet.
Taking advantage of images of the AIA on board SDO, \citet{2016ApJ...828..103T} located many bright plasma blobs within the current sheet during the rise phase of the flare and also interpreted them as plasmoids.
\citet{Gou2019} find that plasmoids moving anti-sunward can merge into a larger one, after coalescence, which eventually causes the formation of a coronal mass ejection.
\citet{2018ApJ...866...64C} propose that the reconnection within the current sheet could be highly turbulent and thus produce plasmoids with different scales. 

Nonthermal broadening of spectral lines detected within the current sheet is another strong indicator of the presence of turbulence \citep[][]{2015SSRv..194..237L}.
Using the Fe \uppercase\expandafter{\romannumeral18} line from the UV Coronagraph Spectrometer (UVCS; \citealt{Kohl1995}) on board the Solar and Heliospheric Observatory (SOHO), \citet{2002ApJ...575.1116C} obtained an upper limit on the nonthermal broadening ($\sim60\,\mathrm{km\,s^{-1}}$) and argued that it is caused by turbulence.
Moreover, \citet{2008ApJ...686.1372C} also obtained the nonthermal broadening of the Fe \uppercase\expandafter{\romannumeral18} line in the current sheet and found that it reaches $380\,\mathrm{km\,s^{-1}}$ at the early stage and decreases to $50$--$200\,\mathrm{km\,s^{-1}}$ later.
\citet{2023FrASS..1096133S} performed a 3D simulation of the flare current sheet reconnection and calculated the synthetic Fe \uppercase\expandafter{\romannumeral21} 1354$\AA$, showing that the turbulent bulk plasma flows contribute to the nonthermal broadening of lines in the reconnection downflows and the loop-top region.
\citet{10.1093/mnras/stae899} propose that nonthermal broadening can also be detected in coronal loops, wherein the turbulence directly originates from the photosphere.

In this study we investigated the observational characteristics of a flare current sheet utilizing data from a high-resolution 3D magnetohydrodynamic (MHD) simulation recently conducted by \citet{2023ApJ...954L..36W}.
This simulation included the necessary physical conditions in the corona and implemented self-sustained turbulent states in both the solar flare current sheet and the loop-top region. 
Synthesizing the remote-sensing data at different wavebands enabled us to determine the relationship between the current sheet's observational characteristics and intrinsic reconnection processes.

This paper is organized as follows. 
The numerical model and simulation results are briefly introduced in Sect.\,\ref{Numerical model}.
We present our analysis results in Sect.\,\ref{Results}, and summarize and discuss our study in Sect.\,\ref{Conclusion and Discussion}.

\begin{figure}
\centering
\resizebox{\hsize}{!}{\includegraphics{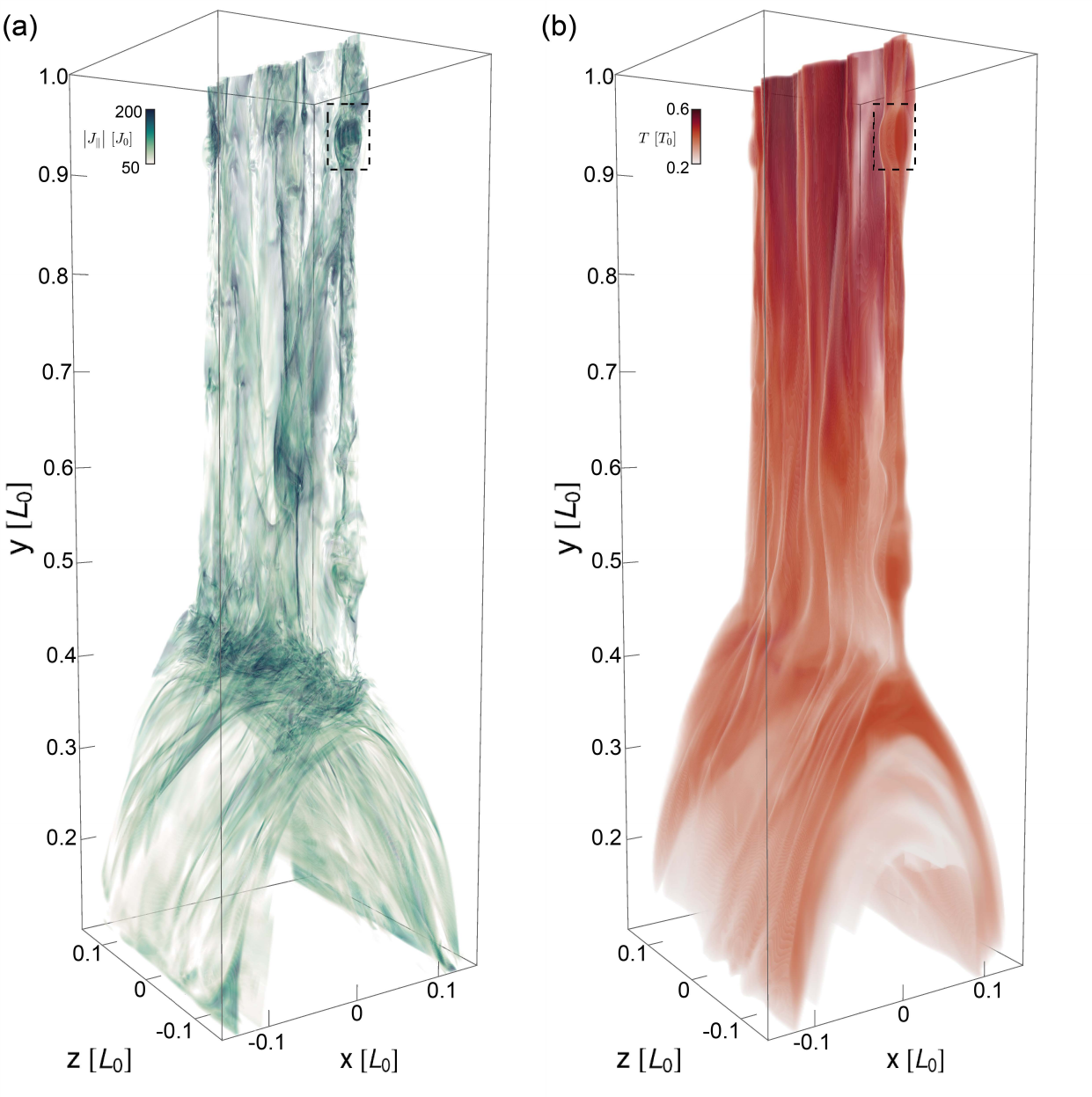}}
\caption{3D structures of parallel current density, $\left|J_\parallel\right|$ (a) and temperature, $T$ (b) at the final moment ($t=8.2$) with fully developed turbulence.
The units of current density and temperature are $J_0=9.54\,\mathrm{statC\,s^{-1}\,cm^-2}$ and $T_0=1.15\times 10^{7}\,\mathrm{K}$, respectively.
The dashed black boxes in both panels highlight a typical flux rope structure.
}
\label{fig:OverviewJT}
\end{figure}

\section{Numerical model}\label{Numerical model}
The simulation by \citet{2023ApJ...954L..36W} solves the resistive MHD equations including the gravity-stratified atmosphere, thermal conduction, radiation cooling, and background heating. 
The simulation domain is set as $x\in\left[-0.5,0.5\right]$, $y\in\left[0,2\right]$, and $z\in\left[-0.15,0.15\right]$.
The unit of length is $L_0=50\,\mathrm{Mm}$.
The initial magnetic field forms a force-free current sheet with a typical CSHKP configuration \citep{1964NASSP..50..451C,1966Natur.211..695S,1974SoPh...34..323H,1976SoPh...50...85K,1995ApJ...451L..83S}.
The current sheet has a length of $100\,\mathrm{Mm}$ and extends about $15\,\mathrm{Mm}$ in the direction of the guide field.
The spatial resolution is uniformly set as $26\,\mathrm{km}$ within the reconnection region ($x\in\left[-0.1,0.1\right]$) to suppress numerical resistivity and obtain the development of turbulence accurately.
The physical time of simulation is $8.2$, where the time unit is $t_0=114\,\mathrm{s}$, the background Alfv\'{e}nic time.

The reconnection is initially triggered by a localized anomalous resistivity $\eta_a=10^{-3}$ at the center of the current sheet, $25\,\mathrm{Mm}$ above the bottom boundary. 
If $\eta_a$ remains constant in time, the resultant evolution will be a standard Petschek-type reconnection \citep{shibata2023}. 
However, in our simulation, it damps temporally and almost vanishes at $t=5$, when the system consists of an erupting principal plasmoid and a thin current sheet comparable with that self-consistently generated by shear motions at the photosphere \citep[see][Fig.\,3]{Dahlin2020}.
The current sheet keeps elongating dynamically until the principal plasmoid moves out of the upper boundary ($y=2$) at $t=6.8$.
Meanwhile, the TMI grows in the current sheet dominated by a background low resistivity $\eta_b$ corresponding to a Lundquist number of $2\times 10^5$, which forms various flux ropes that break later due to the kink instability.
After $t=7.7$, the Kelvin–Helmholtz instability (KHI) is triggered by the sheared flows in the direction of the guide field, forming ray-like structures in the current sheet.
The reconnection in the current sheet finally forms a well-developed turbulent state with an inertial region spanning about one to two magnitude order in wave number space \citep[see][Fig.\,3]{2023ApJ...954L..36W}.

The final current sheet shows a complex turbulent reconnection, manifesting as patchy-like current structures of different scales (see Fig.\,\ref{fig:OverviewJT}a).
Correspondingly, the temperature structure is also highly nonuniform but is spatially smoother compared with the current distribution (Fig.\,\ref{fig:OverviewJT}b), resulting from the quick redistribution of heat by thermal conduction. 
\citet{2023ApJ...954L..36W} has shown that the turbulent state gives rise to many features comparable with observations.
Here we further analyze the data and study the apparent width, heating, and nonthermal velocity of the current sheet. 

\begin{figure}
\centering
\resizebox{\hsize}{!}{\includegraphics{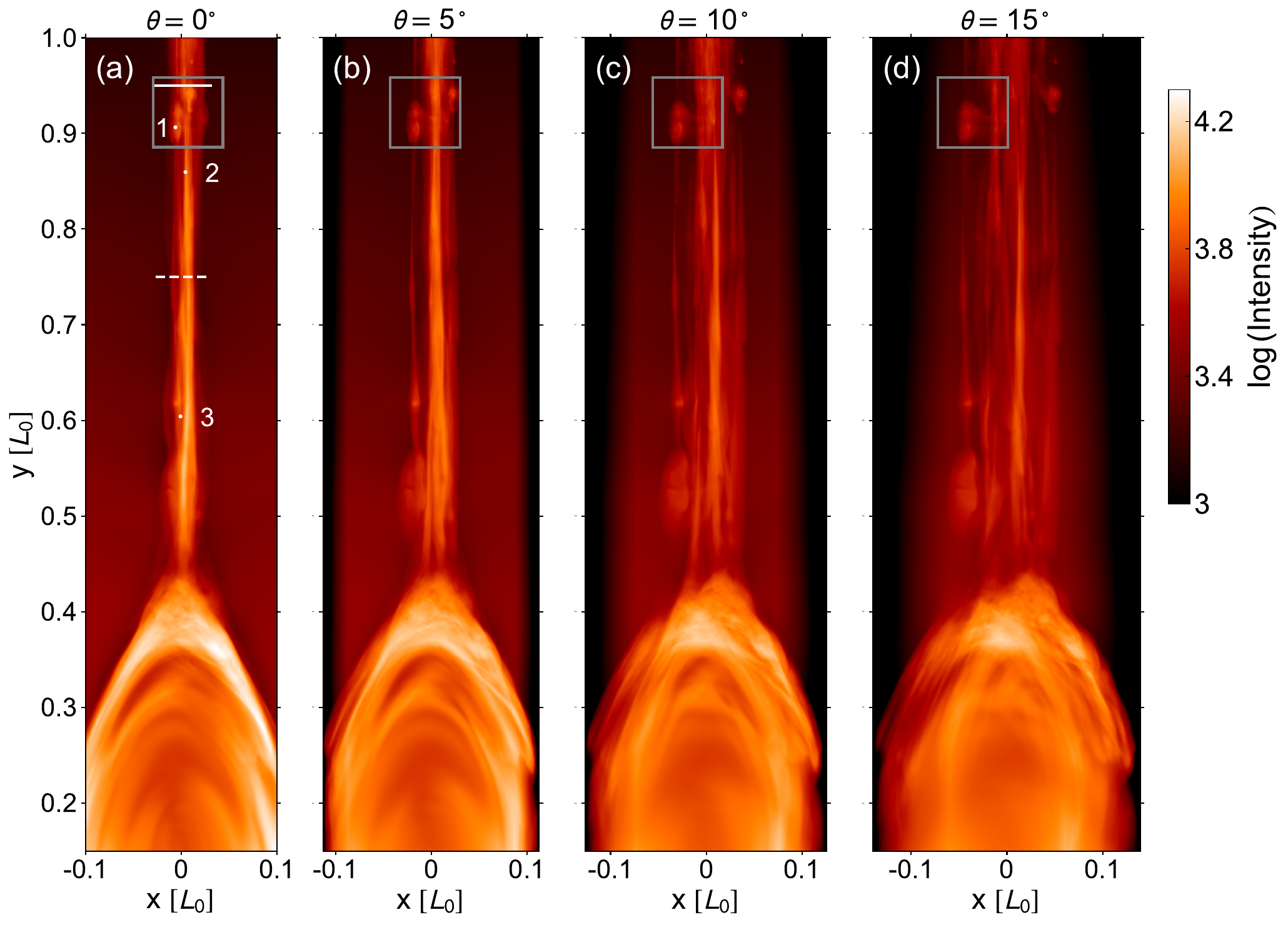}}
\caption{Synthetic X-ray Telescope (XRT) Al-poly/open images of the flare current sheet as observed from different perspectives at $t=8.2$.
Panel (a): Edge-on view, with the LOS in the $z$-direction.
$\theta$ denotes the acute angle between the LOS and the $z$-axis.
The gray boxes in all panels enclose a plasmoid as also highlighted in Fig.\,\ref{fig:OverviewJT}.
The optical thin assumption has been adopted for synthesizing \citep[see][for details]{2023ApJ...954L..36W}.
Panel (a): Two $x$-direction slits used to analyze the apparent and physical widths in Figs.\,\ref{fig:ObsWidthSlit} and \ref{fig:PhysWidthSlice}.\ The three dots mark the ends of three slits along the $z$-direction used in Fig.\,\ref{fig:JTSlits}.
The spatial resolution of XRT is 2 arcsec ($\sim1500\,\mathrm{km}$; \citet{XRT}), corresponding to $0.03L_0$ in the simulation.
}
\label{fig:XRT_MVP}%
\end{figure}

\begin{figure}
\centering
\resizebox{\hsize}{!}{\includegraphics{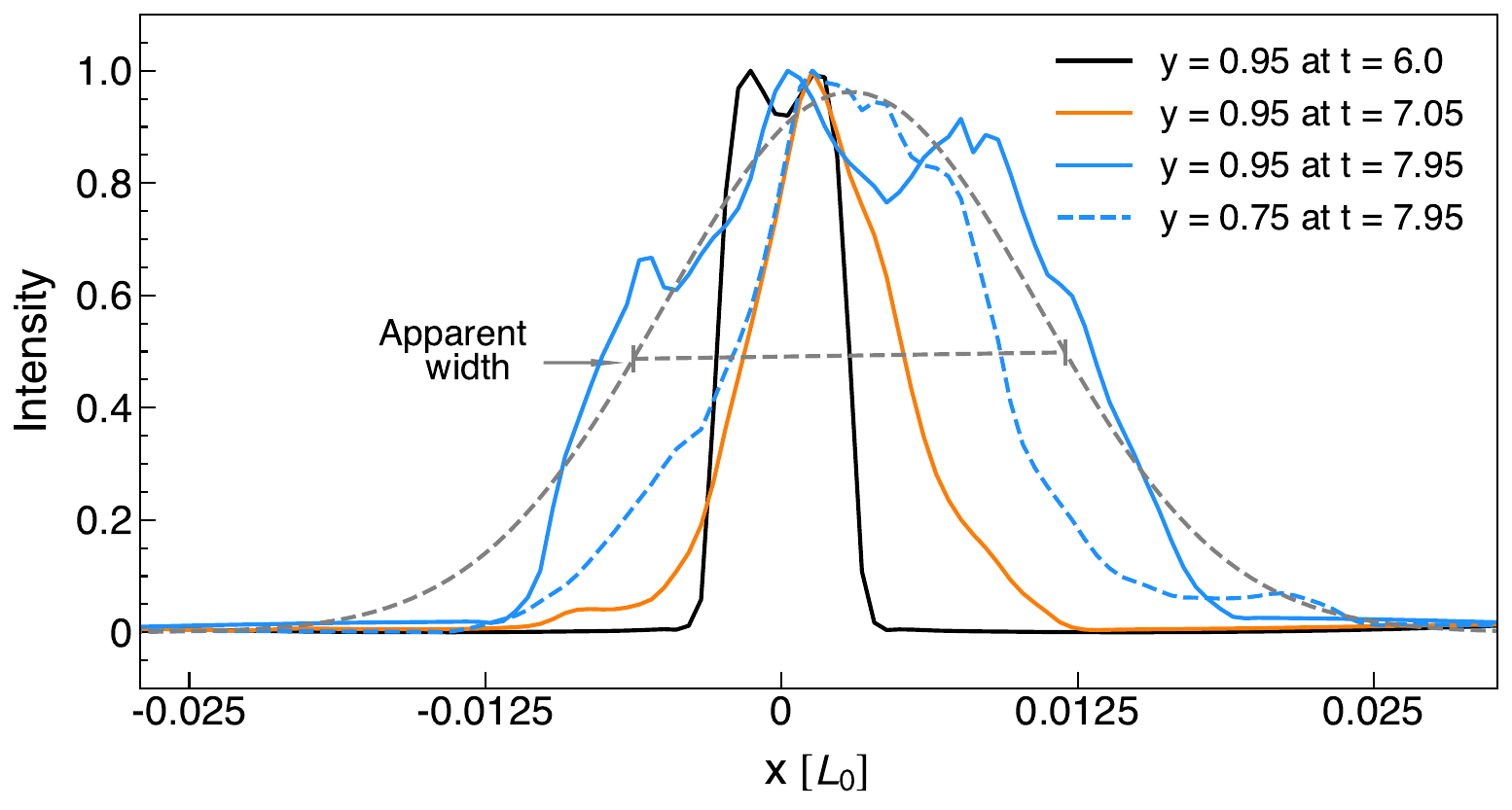}}
\caption{Profiles of synthetic XRT intensity across the current sheet at different moments.
The solid curves with different colors depict the intensity profiles at three typical moments, namely $t=6$, $7.05$, and $7.95$, sampled along the $x$-direction slit at $y=0.95$ (see the solid white line in Fig.\,\ref{fig:XRT_MVP}a). 
The dashed blue curve shows the result of another slit ($y=0.75$) at $t=7.95$ (see the dashed white line in Fig.\,\ref{fig:XRT_MVP}a).
The dashed gray curve denotes the single Gaussian fitted to the solid blue curve, on which the definition of the apparent width is marked.
All these curves have been normalized by their maximum value for a better comparison.
}
\label{fig:ObsWidthSlit}%
\end{figure}

\begin{figure*}
\centering
\includegraphics[width=0.8\textwidth]{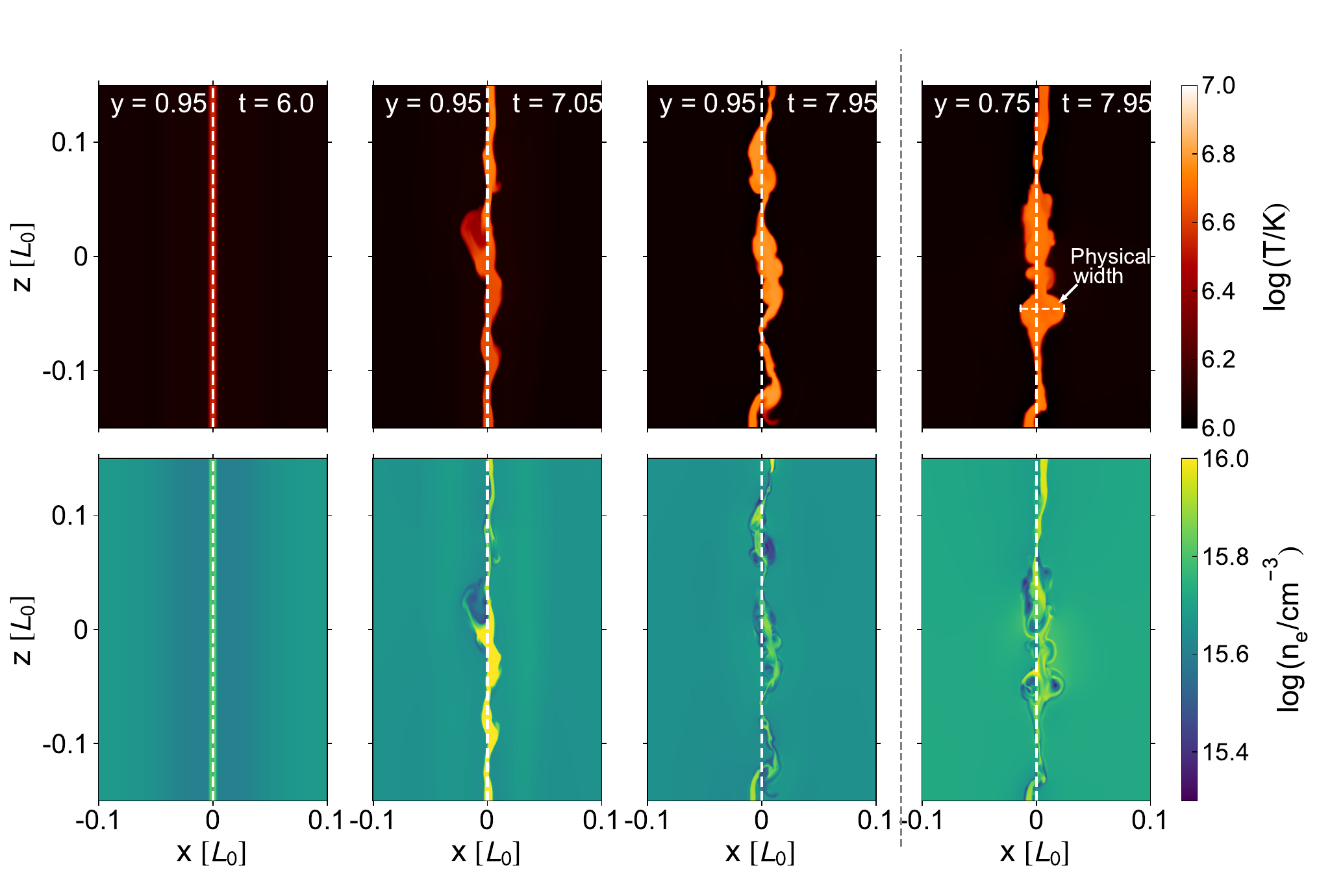}
\caption{Horizontal slices of temperature (first row) and density (second row) at typical moments, the same as in Fig.\,\ref{fig:ObsWidthSlit}.
The first three columns depict the slices of $y=0.95$ at $t=6$, $7.05$, and $7.95$, and the fourth column plots the slice of $y=0.75$ at $t=7.95$ on which the definition of the physical width is denoted.
The dashed white lines mark the position of $x=0$.
}
\label{fig:PhysWidthSlice}%
\end{figure*} 

\begin{figure}
\centering
\resizebox{\hsize}{!}{\includegraphics{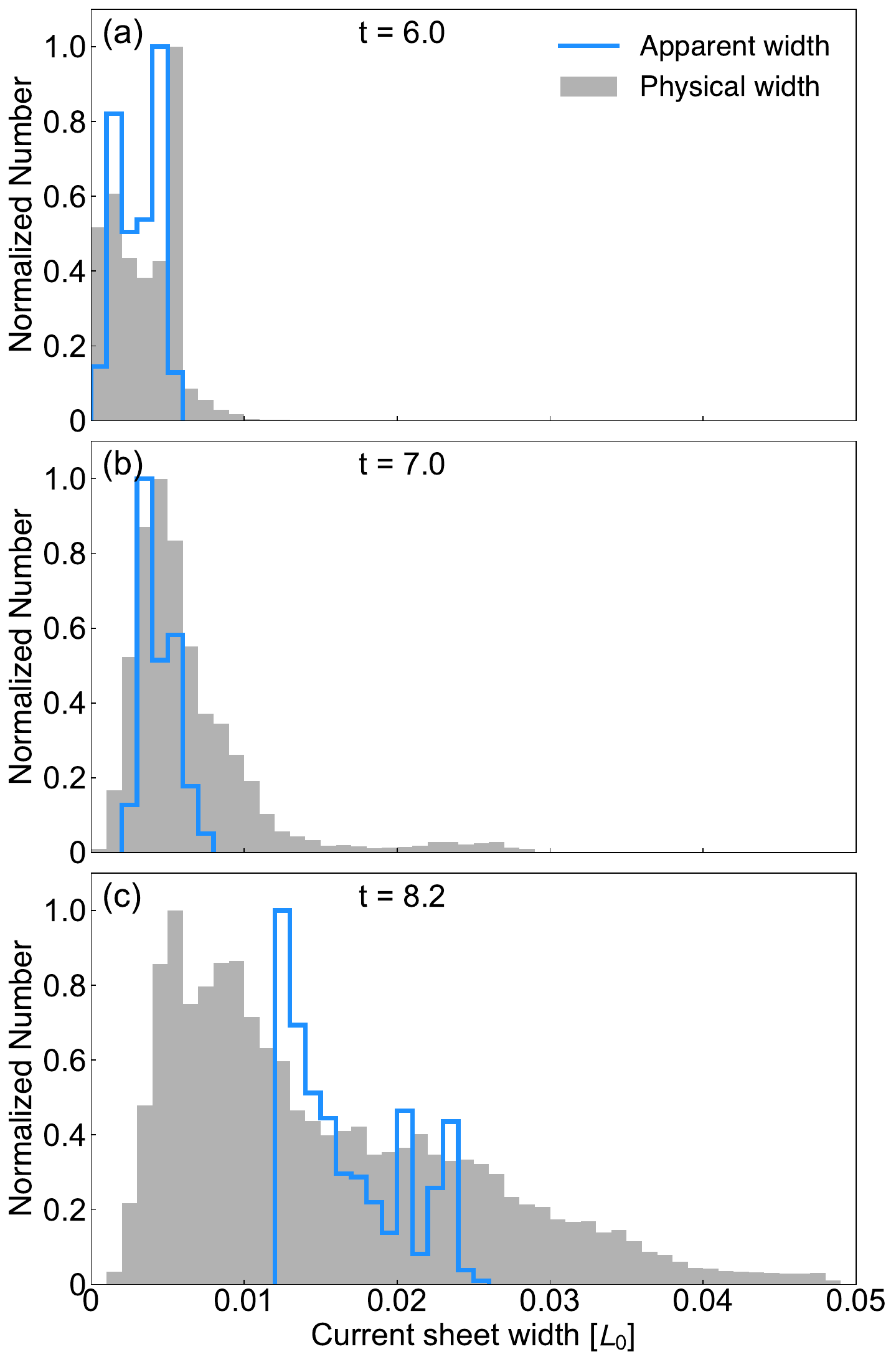}}
\caption{Statistics of the apparent and physical width of the current sheet at different evolution stages.  
The blue curves depict the distributions of the apparent width, and the gray shades exhibit the histograms of the physical width.
The bin size is $52\,\mathrm{km}$.
}
\label{fig:WidthHist}%
\end{figure}

\section{Results}\label{Results}

\subsection{The current sheet width}

\subsubsection{The influence of 3D effects}
In observations, the widths of current sheets are frequently estimated by the transverse length observed from the edge-on view \citep{2018ApJ...866...64C}.
However, the superposition effect makes such a width an apparent value. Generally speaking, the realistic current sheet is 3D and will present different structures as observed from other directions.
As shown by Fig.\,\ref{fig:XRT_MVP}, as the line-of-sight (LOS) direction deviates from the edge-on perspective, the manifestation of the current sheet will change significantly.
To be specific, its observed width largely changes, accompanied by the appearance of fine structures.
If the telescope cannot resolve the small-scale bright structures inside the current sheet, the apparent width will increase significantly as deviating from the edge-on direction (compare Fig.\,\ref{fig:XRT_MVP}a and d).
Therefore, the 3D effects do influence the determination of the current sheet width from remote-sensing data.
We note that some relatively small-scale brighter structures, which correspond to 3D flux ropes with strong current and high temperature (see the dashed black boxes in Fig.\,\ref{fig:OverviewJT}), can always be recognized independent of the direction (see the gray boxes in Fig.\,\ref{fig:XRT_MVP}).

\subsubsection{The evolution of the current sheet width}\label{width}
To simplify the analysis and also compare with the observations of SOL2017-09-10T X8.2 flare \citep[][]{2018ApJ...866...64C}, hereafter we use the synthetic images from the edge-on direction to represent the main features of the apparent width of the current sheet (Fig.\,\ref{fig:XRT_MVP}a).
We first selected a horizontal slit across the current sheet at $y=0.95$ (the solid white line in Fig.\,\ref{fig:XRT_MVP}a) and extracted the synthetic XRT intensity profiles at different moments to investigate the evolution of the apparent width.
As shown by Fig.\,\ref{fig:ObsWidthSlit}, at the initial stage, $t=6$, the XRT intensity distribution is almost symmetric with regard to the $x=0$ plane (the black curve in Fig.\,\ref{fig:ObsWidthSlit}).
At this moment, the current sheet still tends to exhibit a quasi-2D structure (see the first column of Fig.\,\ref{fig:PhysWidthSlice}).  
As reconnection develops, the intensity profile gradually widens (the orange curve in Fig.\,\ref{fig:ObsWidthSlit}).
This is due to the formation of flux ropes by TMI, thus causing the expansion of the current sheet at corresponding positions (see also the second column of Fig.\,\ref{fig:PhysWidthSlice}).
In particular, as the KHI starts, various vertically distributed flux ropes are formed as threaded by curled field lines \citep[see][]{2023ApJ...954L..36W}.
The current sheet also gets distorted in the $z$-direction, especially at relatively thin regions (see the third and fourth columns of Fig.\,\ref{fig:PhysWidthSlice}). 
Hence, the apparent width of the current sheet further increases as viewed along the LOS (see the solid blue curve in Fig.\,\ref{fig:ObsWidthSlit}).
Accompanied by the broadening of the current sheet width, the shape of the XRT intensity curve also gets complex and represents multiple peaks that correspond to small-scale structures formed inside the current sheet.

\subsubsection{Comparison of apparent and physical widths}
To quantitatively investigate the differences between the observed and real current sheet width, we defined two types of widths, namely the apparent and physical widths.
The apparent width at a height $y_i$ is defined as the FWHM of the single Gaussian fitted function of the XRT intensity profile along the horizontal slit at $y=y_i$ (see Fig.\,\ref{fig:ObsWidthSlit}), where the subscript ''$i$'' refers to the $i$th grid in the $y$-direction.
We used single Gaussian fitting to eliminate the details in the XRT intensity profiles.
This definition basically follows up previous observational methods \citep[see][]{2003ApJ...594.1068K,2018ApJ...866...64C}.

The physical width of the current sheet at a position $\left(y_i, z_j\right)$ is defined as the temperature width in the $x$-direction, where the subscript ``$j$'' refers to the $j$th grid point in the $z$-direction.
The temperature width is determined as the distance between the positions with extremal temperature gradients at both outer edges of the current sheet (see the last column of Fig.\,\ref{fig:PhysWidthSlice}).
Because of the thermal conduction, the temperature differences along the magnetic field inside the current sheet can be quickly smoothed out.
However, as the thermal conduction perpendicular to field lines is negligible, the internal temperature is significantly higher than the external temperature, creating a large temperature change near the boundaries of the current sheet.
As a result, the temperature width can approximately represent the width of the reconnection region \citep[see also][]{Wang2022}. 
Our definition of the physical width is similar to the outer scale of the reconnection region proposed by \citet{2022ApJ...940...94B}.
According to this definition, the physical width has a 2D distribution on the $y$-$z$ plane, while the apparent width is distributed only along the $y$-direction.

We counted all apparent and physical widths of the current sheet region defined by $y\in\left(0.47,1\right)$ and $z\in\left(-0.15,0.15\right)$ and compare their statistical distributions in Fig.\,\ref{fig:WidthHist}.
During the initial stage ($t=6$), the two types of widths have similar profiles and concentrate at a narrow region with small values (Fig.\,\ref{fig:WidthHist}a), consistent with the globally thin current sheet before the growth of TMI (see also the first column of Fig.\,\ref{fig:PhysWidthSlice}).
Later, the distributions of the apparent and physical widths start to separate.
We first compared the physical widths at $t=7$ and $t=8.2$ (Figs.\,\ref{fig:WidthHist}b and c), which correspond to the moments before and after the turbulence is fully developed, respectively.
At $t=7$, the physical widths are mainly smaller than 0.01, while a small proportion can reach much larger values related to several relatively large flux ropes produced by the TMI (see the tail of the gray histogram in Fig.\,\ref{fig:WidthHist}b).
At $t=8.2$, though the peak of the physical width is still located near 0.01, a considerable number of samples with greater widths appear and cause the extension of the profile tail (see the gray shade in Fig.\,\ref{fig:WidthHist}c).
They reflect the emerging vortex structures produced by the turbulence and KHI.

Then we focused on the variation of the apparent width from $t=7$ to $8.2$.
The peak of the apparent width moves to a larger value after turbulence is well developed at $t=8.2$ (Fig.\,\ref{fig:WidthHist}b and c).
At both moments, a few current sheet regions grow to form the larger current patches, resulting in larger physical widths, which, however, cannot be resolved by the observations (see the gray shades on the right of the blue curves in Fig.\,\ref{fig:WidthHist}b and c).
The reason is that the FWHM of LOS-integrated images mainly resolves the brightest central region, while the darker halos on both sides contributed by a few wider structures are not counted (see the second column of Fig.\,\ref{fig:PhysWidthSlice}).  
More interestingly, the current sheet regions with physical widths smaller than 0.01 constitute the peaks at both moments, which can be recognized by the apparent width at $t=7$ (Fig.\,\ref{fig:WidthHist}b) but missed at $t=8.2$ (Fig.\,\ref{fig:WidthHist}c).
At $t=7$, the current sheet shows a relatively laminar pattern, so the majority of current patches with small widths are still spatially connected, forming a globally thin sheet that can be well resolved by observations.
However, once the turbulence is fully developed, despite many current patches having small widths, they have chaotic distributions and orientations and are thus apparently connected to each other and form larger structures as seen from the edge-on view.
Therefore, the apparent width is much larger than the realistic physical width of the current sheet region.

\begin{figure}
\centering
\resizebox{\hsize}{!}{\includegraphics{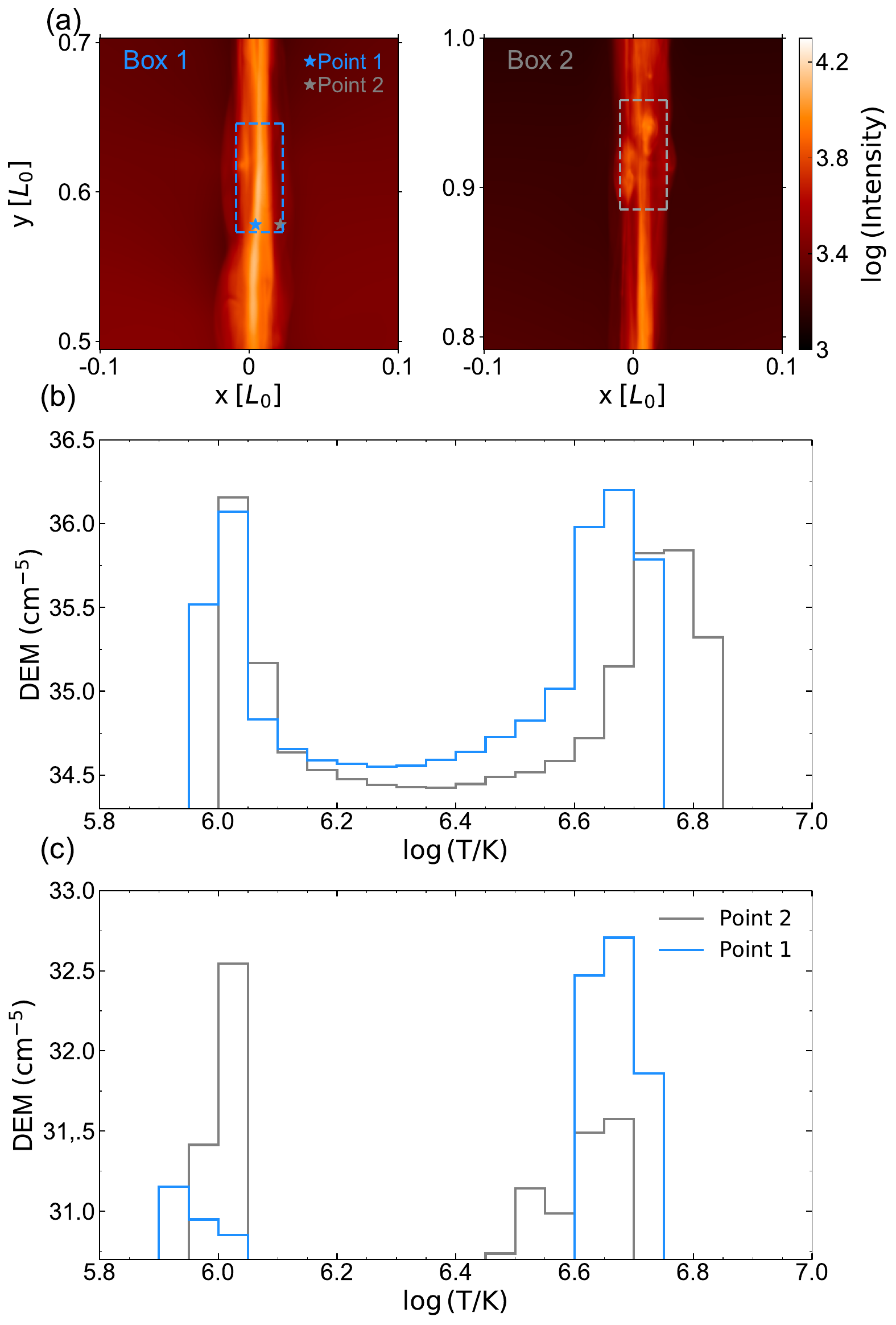}}
\caption{DEM of the current sheet at two specific regions.
Panel (a): Two zoomed-in regions of the XRT image of the current sheet at $t=8.2$.
The left region encloses a ray-like segment, and the right one contains several plasmoids.
Panel (b): DEMs of box 1 and box 2, labeled in panel (a).
The DEM is derived following the method proposed by \citet{2012ApJS..203...25G}.
We collect the temperature information of grid points within the boxes for statistics; the bin size of DEMs is $log\,\mathrm{T}=0.05$.
Panel (c): DEMs of two points, marked by stars in panel (a).
}
\label{fig:EM}%
\end{figure}

\begin{figure}
\centering
\resizebox{\hsize}{!}{\includegraphics{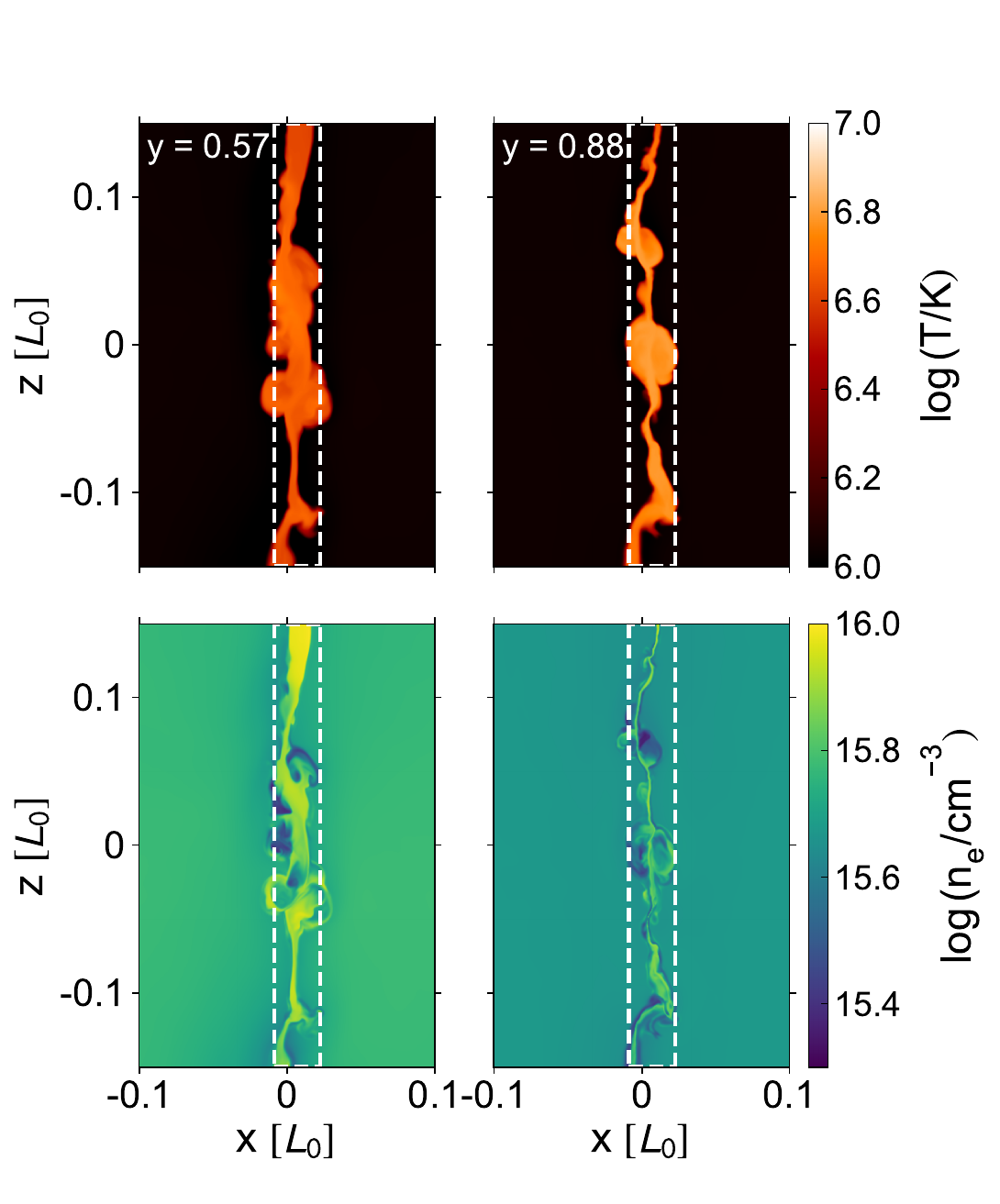}}
\caption{Same horizontal slices of temperature and density
as in Fig.\,\ref{fig:PhysWidthSlice} but at $t=8.2$ and in boxes 1 and 2 (labeled in Fig.\,\ref{fig:EM}a). The dashed white line indicates the box we selected to calculate DEMs.}
\label{fig:8.2slice}%
\end{figure}

\begin{figure}
\centering
\resizebox{\hsize}{!}{\includegraphics{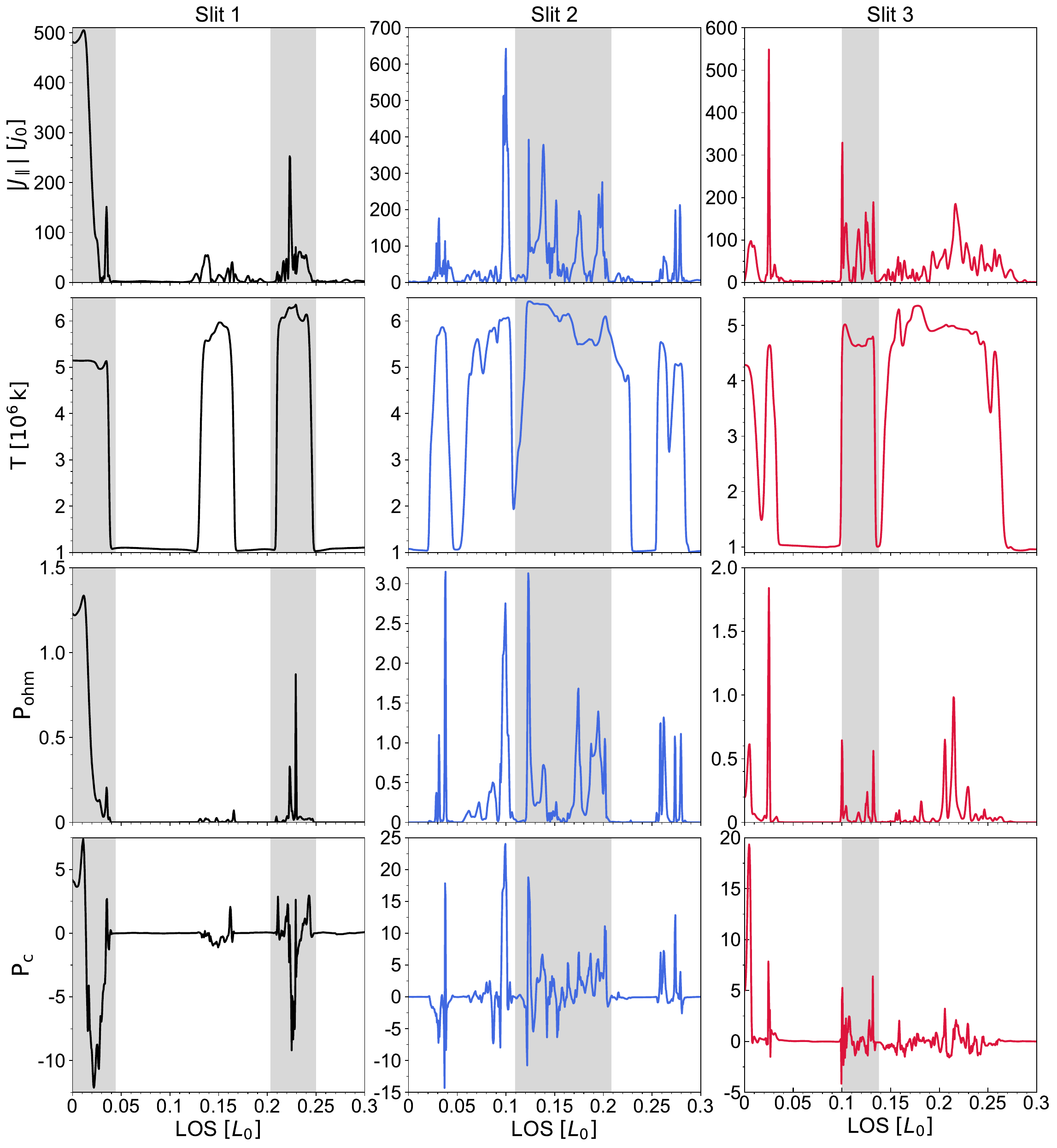}}
\caption{LOS profiles of $\left|J_\parallel\right|$, $\mathrm{T}$, $P_{ohm}$, and $P_c$ at three typical locations, labeled by white dots in Fig.\,\ref{fig:XRT_MVP}a.
The gray shades highlight several regions with high temperatures and large parallel currents.
The unit of power density is $P_0=\mathrm{0.28\,erg\cdot cm^{-3}\cdot s^{-1}}$.
}
\label{fig:JTSlits}%
\end{figure} 
    
\subsection{DEM of the current sheet}
A commonly used technique in solar observations is the DEM, which can provide the emission measure distributions as a function of temperature \citep{2012ApJS..203...25G,chengetal.2012}.
In this section, we focus on typical observational structures in the synthetic XRT images, and analyze their DEMs and relations with the fine structures of turbulent reconnection \citep[see][]{2018ApJ...866...64C}.

In Fig.\,\ref{fig:EM}a, box 1 encloses a bright structure near the center of the current sheet.
The DEM of the area has two peaks (see the blue curve in Fig.\,\ref{fig:EM}b).
The peak at $10^6\,\mathrm{K}$ corresponds to the background coronal temperature we set in the simulation, while the high-temperature peak is related to the plasmas heated by the reconnection inside the current sheet, indicating the plasmas over there have a high temperature.
Box 2 encloses several bright structures that correspond to plasmoids in observations \citep{2016ApJ...828..103T}.
Its DEM is similar to that of box 1 and also presents two peaks but the right peak has a higher temperature (see the gray curve in Fig.\,\ref{fig:EM}b).
In both cases, the low-temperature peaks are comparable with the high-temperature ones in magnitude, implying that the boxes we selected contain enough low-temperature components surrounding the highly deformed current sheet (see also Fig.\,\ref{fig:8.2slice}). 
To illustrate the differences in the DEM distribution at different regions, we show the results at two points within box 1 (see Fig.\,\ref{fig:EM}a and c).
Point 1 is located at the center of the current sheet and its DEM primarily exhibits one high-temperature component at $log\,\mathrm{T}=6.7$ (see the blue curve of Fig.\,\ref{fig:EM}c). 
In contrast, point 2 is set at the edge of the current sheet, which presents a low-temperature peak even higher than the high-temperature one (see the gray curve of Fig.\,\ref{fig:EM}c).

We next investigated the origin of the high-temperature components.
In Fig.\,\ref{fig:JTSlits}, we compare the LOS distributions of $\left|J_\parallel\right|$, $\mathrm{T}$, Ohmic heating power $P_{ohm}=\eta_b J^2$, and compression heating power $P_c=-p\nabla \cdot \boldsymbol{u}$ at three typical positions.
Point 1 is selected inside the plasmoid structure, point 2 is set in a bright structure, and point 3 is located at a relatively dark region near a bright structure  (see Fig.\,\ref{fig:XRT_MVP}a).
For a uniform $\eta_b$, the distribution of strong $J_\parallel$ can reflect the 3D reconnection sites \citep[][]{Reid2020,Wang2024}.
According to Fig.\,\ref{fig:JTSlits}, though the temperature profiles are smoothed by thermal conduction, all of the high-temperature regions correspond to those with strong parallel currents, which implies that the high-temperature components of the DEMs are approximately related to the reconnection sites distributed along the LOS.
As the two local heating sources, the regions with strong $P_{ohm}$ and $P_c$ are also spatially related to reconnection sites (see the third and fourth rows of Fig.\,\ref{fig:JTSlits}).
In regions with $\left|J_\parallel\right|>10$, the total Ohmic heating power is $P_{ohm}=3.04\times10^6$ about 14 times that in other regions.
Similarly, the compression heating power in $\left|J_\parallel\right|>10$ regions is $P_c=6\times10^6$ even larger than $P_{ohm}$, which, however, becomes $-4.04\times10^5$ in the other regions.
It should also be noted that, besides the regions labeled by gray shades in Fig.\,\ref{fig:JTSlits}, there are several other high-temperature segments but with relatively weak $J_\parallel$ and local heating.
This phenomenon implies that, under a strong turbulent state, besides local heating, thermal conduction can also transfer energies from remote locations along the chaotic magnetic field lines connecting multiple reconnection sites.

\begin{figure}
\centering
\resizebox{\hsize}{!}{\includegraphics{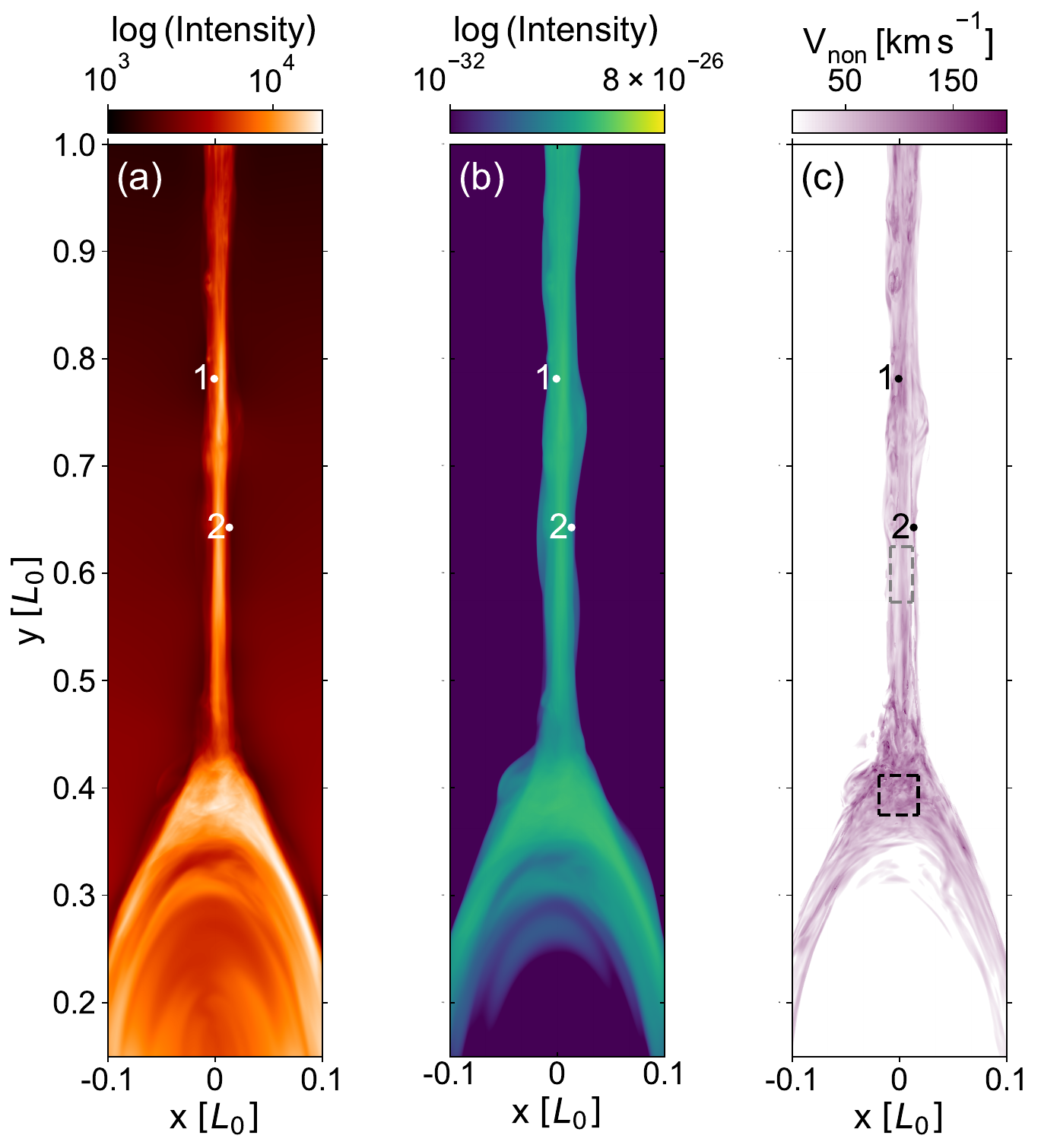}}
\caption{Synthetic images of XRT Al-poly/open (a), Ca\uppercase\expandafter{\romannumeral17} line (b), and nonthermal velocity (c) observed from the edge-on view at $t=8$.
The dots labeled 1 and 2 mark the two end-of-slit positions used to produce a synthetic line profile in Fig.\,\ref{fig:CaXVII}.
The dashed black and gray box in panel (c) shows the region we selected to calculate the nonthermal velocity in Fig.\,\ref{fig:timevarynonv}.}
\label{fig:ObsNonThermal}
\end{figure}

\begin{figure}
\centering
\resizebox{\hsize}{!}{\includegraphics{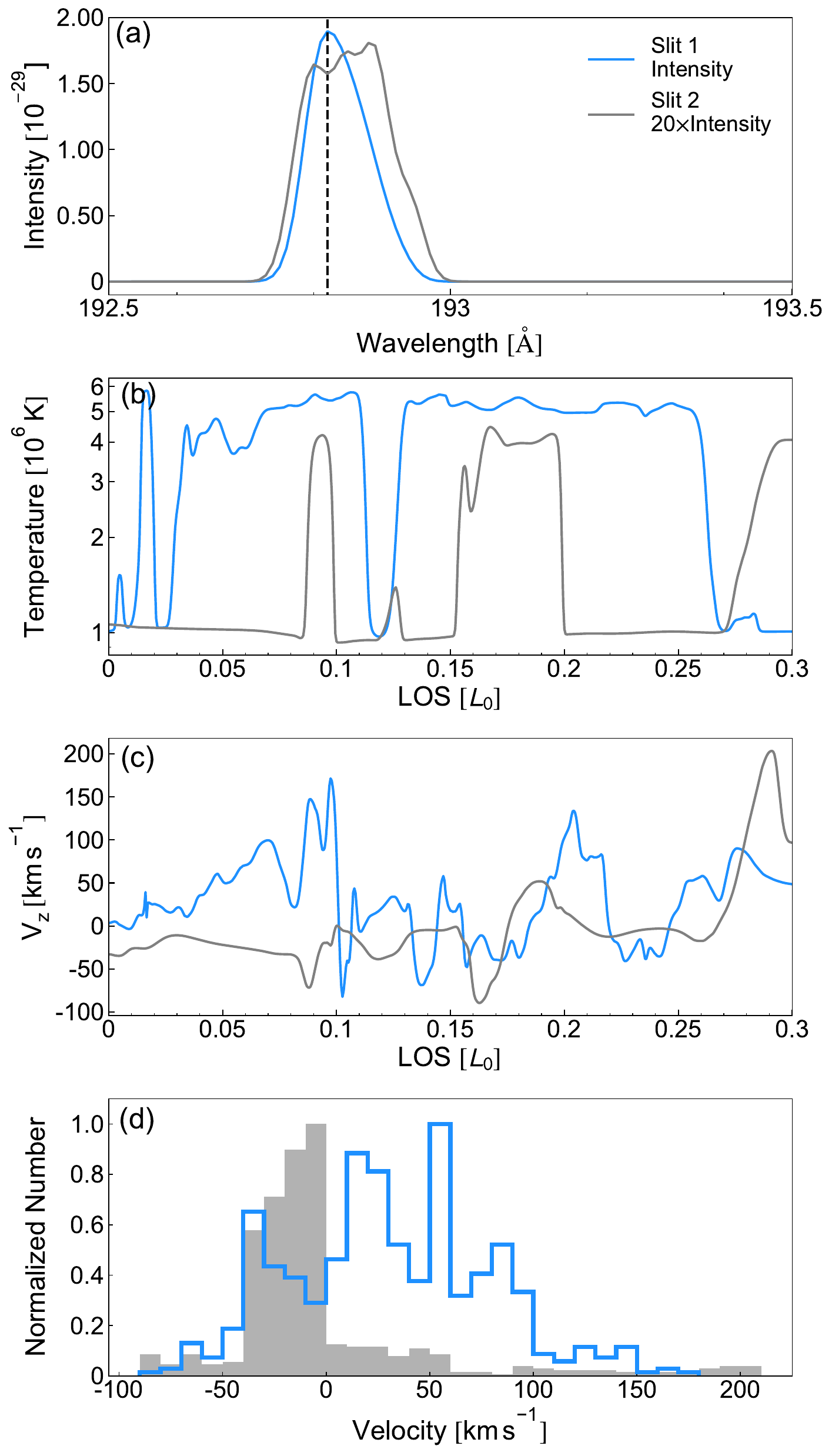}}
\caption{Synthetic Ca\uppercase\expandafter{\romannumeral17} emission lines (a), LOS profiles of temperature (b), and $V_z$ (c) at two typical positions (labeled 1 and 2 in Fig.\,\ref{fig:ObsNonThermal}) at $t=8$.
The spectrum lines in panel (a) were obtained by integrating the local spectra along the LOS.
Panel (d) shows the histograms of the $V_z$ sampled along the LOS.
}
\label{fig:CaXVII}%
\end{figure} 

\begin{figure}
\centering
\resizebox{\hsize}{!}{\includegraphics{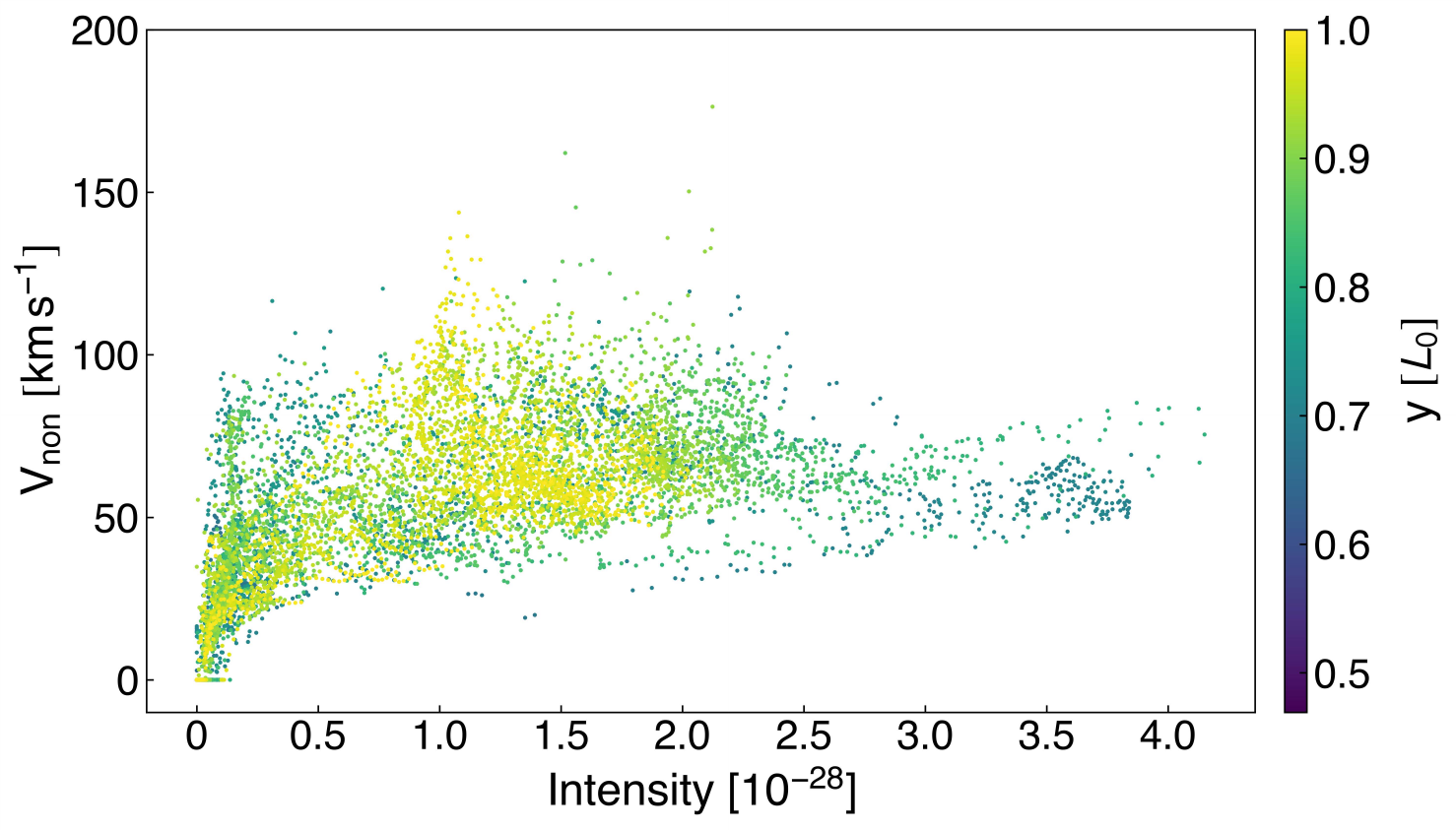}}
\caption{Relation between the nonthermal velocity and the synthetic spectral intensity.
Different colors indicate the different heights.
}
\label{fig:NTHist}%
\end{figure} 

\begin{figure}
\centering
\resizebox{\hsize}{!}{\includegraphics{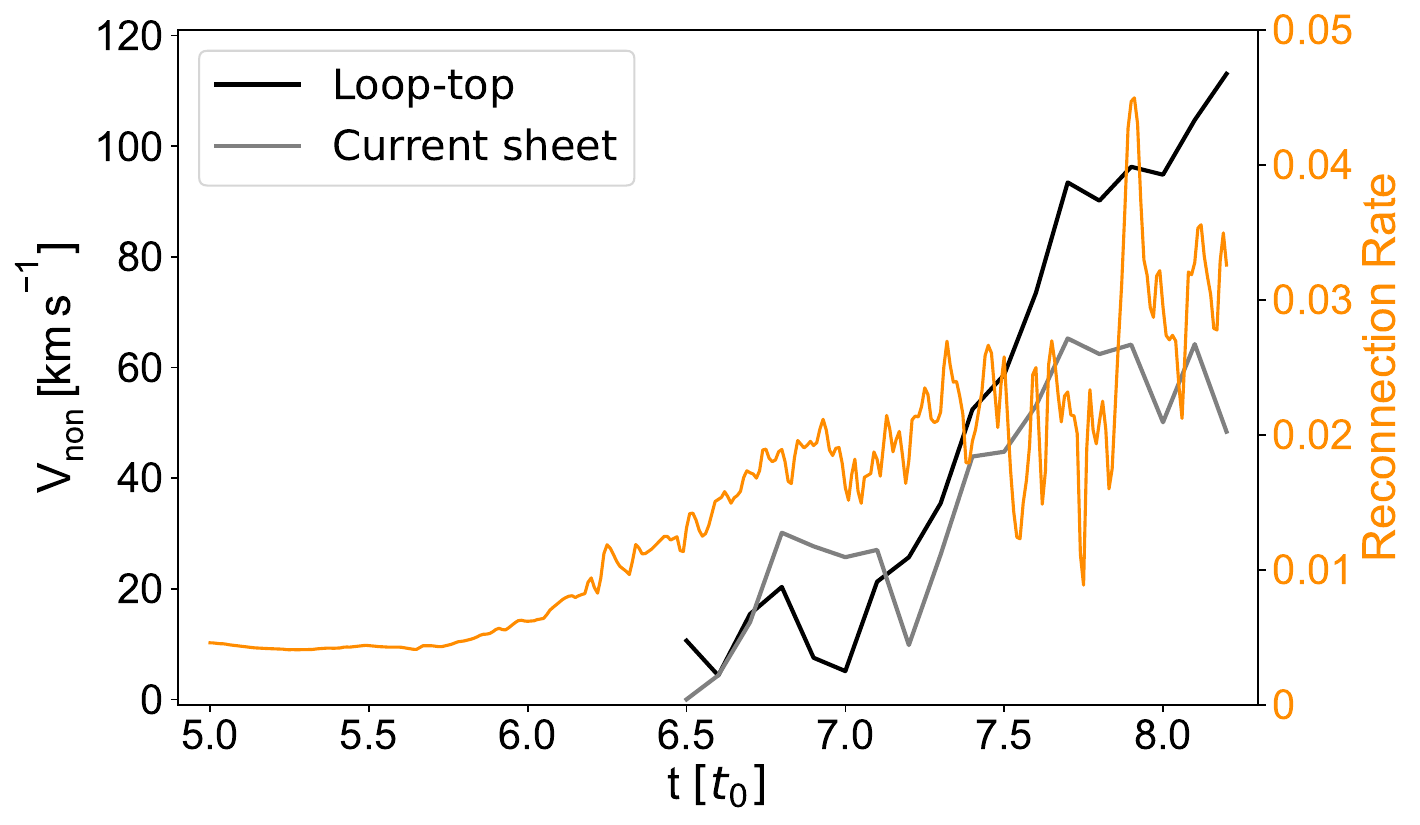}}
\caption{Time evolution of the nonthermal velocity in the current sheet and loop-top region (marked by the dashed gray and black box in Fig.\,\ref{fig:ObsNonThermal}), as well as the reconnection rate (orange line).
}
\label{fig:timevarynonv}%
\end{figure} 

\subsection{Nonthermal broadening of spectral lines}

In our simulation, the turbulence has been well developed both in the current sheet and the loop-top regions.
Therefore, there exist various velocity perturbations in the reconnection regions, which, especially the motions along the LOS, can potentially cause the nonthermal broadening of spectral lines. 
Limited by the maximal temperature in the simulation, we synthesized the Ca \uppercase\expandafter{\romannumeral17} line, which is available from the EUV Imaging Spectrometer (EIS) on the Hinode satellite  \citep{2007SoPh..243...19C}.
We chose this spectral line because its typical temperature is $10^{6.7}\,\mathrm{K}$, which corresponds to the main high-temperature components in our simulation (see Fig.\,\ref{fig:EM}b). 
Thus, the turbulent flows caused by reconnection can be captured.

We synthesized the emission line Ca \uppercase\expandafter{\romannumeral17} as observed from the edge-on direction following the method outlined in \citet{2017ApJ...846L..12G} and \citet{2023FrASS..1096133S}: the integrated intensity of Ca 
\uppercase\expandafter{\romannumeral17} at a frequency $\nu$ was calculated as
\begin{equation}
I_{\nu} = \frac{h\nu}{4\pi}\int f_{\nu}n_{e}n_{H}g(T_{e})dl\,,\label{eq:Inu} 
\end{equation}
where $h$ is the Planck constant, $n_e$ and $n_H$ respectively denote the number densities of electrons and protons, $g(T_e)$ is the contribution function available at the CHIANTI database \citep{2021ApJ...909...38D}, and 
\begin{equation}
f_\nu = \frac1{\pi^{1/2}\Delta\nu_D}\exp(-(\frac{\Delta\nu+\nu_0v_l/c}{\Delta\nu_D})^2)\,\label{eq:fnu}
\end{equation}
calculates the local spectral line profile of plasma moving in the LOS direction but in a local thermal equilibrium state, where $\nu_0$ is the rest frequency of Ca \uppercase\expandafter{\romannumeral17}, $c$ is the speed of light in vacuum, $v_l$ is the plasma velocity along the LOS, $\Delta\nu$ denotes the frequency offset with regard to the central frequency, and $\Delta\nu_D = \nu_0\sqrt{2kT/m}/c$ is the thermal broadening in the frequency domain.

Figure \ref{fig:ObsNonThermal}b provides a synthetic Ca \uppercase\expandafter{\romannumeral17} image of the current sheet at $t=8$, which is obtained through integrating the spectral intensity over the wavelength range ($191\AA$ -- $195\AA$) (see Fig.\,\ref{fig:ObsNonThermal}b).
Similar to the XRT image (Fig.\,\ref{fig:ObsNonThermal}a), the Ca \uppercase\expandafter{\romannumeral17} image also clearly captures the current sheet and the loop top. 
The difference is that the Ca \uppercase\expandafter{\romannumeral17} image exhibits a wider current sheet (see Figs.\,\ref{fig:ObsNonThermal}a and b), mainly because the two channels ($log\,\mathrm{T}=6.7$ for Ca\uppercase\expandafter{\romannumeral17} and $log\,\mathrm{T}=6.9$ for XRT) have different typical temperatures.
Moreover, according to Fig.\,\ref{fig:EM}b, temperatures within the central current sheet can exceed $10^{6.7}$ K.
Therefore, the Ca\uppercase\expandafter{\romannumeral17} image misses the highest-temperature components and instead includes more contributions from lower-temperature plasma, namely the halo around the current sheet.

We selected two typical positions, 1 (near the center) and 2 (near the edge), to analyze their spectral lines (see Fig.\,\ref{fig:ObsNonThermal}).
The synthetic line profiles at both positions show similar red-shift features but also have two main differences (see Fig.\,\ref{fig:CaXVII}a).
First, the intensity at position 1 is much stronger than that at point 2, mainly because the average temperature at point 1 is closer to the typical temperature of the Ca \uppercase\expandafter{\romannumeral17} line ($log\,\mathrm{T}=6.7$).
Second, the line at position 1 shows a single Gaussian profile, while that at position 2 exhibits a complex multi-peak profile.
This difference mostly originates from their different velocities and temperatures along the LOS.
On the one hand, the LOS velocity at point 1 along LOS oscillates rapidly around zero compared with that at point 2 (Fig.\,\ref{fig:CaXVII}c), corresponding to a wider velocity dispersion (see Fig.\,\ref{fig:CaXVII}d).
On the other hand, according to Eq.\,\ref{eq:Inu}, the Ca \uppercase\expandafter{\romannumeral17} line is only sensitive to high-temperature plasma.
For point 1, $80\%$ of the regions along the LOS are of high temperature, while for 
point 2, only a few regions have a temperature where Ca \uppercase\expandafter{\romannumeral17} line is sensitive (Fig.\,\ref{fig:CaXVII}b).

Similar to the method used in \citet{2018ApJ...854..122W}, we also estimated the nonthermal velocity caused by turbulence using the formula
\begin{equation}
\mathrm{FWHM}=\sqrt{4\ln2{\left(\frac{2kT_{eq}}m+v_{nth}^2\right)}}\frac{\lambda_0}c\,,
\label{eq:nthV}
\end{equation}
where FWHM is that of the spectral line and $T_{eq}$ is the typical forming temperature of Ca \uppercase\expandafter{\romannumeral17} line.
We calculated the FWHM via the single Gaussian fit, while for the line profiles with multiple peaks, we applied a single Gaussian fit to each peak and calculated the FWHM as the square root of the sum of the squares of all FWHMs.
The nonthermal velocities at positions 1 and 2 in Fig.\,\ref{fig:ObsNonThermal} given by this method are $85\,\mathrm{km\,s^{-1}}$ and $200\,\mathrm{km\,s^{-1}}$, respectively, which are similar to typical values in observations \citep[see][]{2018ApJ...853L..15L,2014ApJ...797L..14T}. 
In Fig\,.\ref{fig:ObsNonThermal}c, we present the nonthermal velocity map at $t=8$, the moment where the turbulent reconnection is fully developed.
It shows that the nonthermal velocity can exceed $100\,\mathrm{km\,s^{-1}}$ in the loop-top region, and can also reach several tens of $\mathrm{km\,s^{-1}}$ in the current region.

We next investigated the relation between the nonthermal velocity and the integrated line intensity, which has been frequently discussed in previous observational works \citep{2014ApJ...788...26D,2018ApJ...853L..15L}.
We focused on the region of $y>0.47$ (the current sheet region), where the nonthermal velocity mainly originates from the self-sustained turbulence compared with the loop-top region.
In Fig. \ref{fig:NTHist} we present the distribution of the nonthermal velocity relative to the line intensity and use different colors to represent sample points with different heights.
According to Fig.\,\ref{fig:NTHist}, there is no clear correlation between the nonthermal intensity and the spectral intensity, which can also be qualitatively reflected by comparing Fig\,.\ref{fig:ObsNonThermal}b and c.
Most of the nonthermal velocities are distributed at the range of $50$--$100\,\mathrm{km\,s^{-1}}$, almost independent of the intensities.
Moreover, the heights also have little effect on the nonthermal velocities, which implies that the turbulence is almost uniformly distributed in the current sheet.

Finally, in Fig.\,\ref{fig:timevarynonv} we compare the temporal variation of the nonthermal velocity with the development of the reconnection rate and turbulence. 
We selected two boxes at the current sheet and loop-top regions (see Fig.\,\ref{fig:ObsNonThermal}c) and calculated the average nonthermal velocity therein.
To be specific, we summed the profiles at all grid points within a box to form an average profile, from which we then calculated the average nonthermal velocity.
Before $t=6.5$ (the initial stage of fast reconnection), the average nonthermal velocities in both regions are ignorable.
Later, the nonthermal velocities increase significantly, temporally consistent with the increase in the reconnection rate and turbulence. 
It should also be noted that for many flares the timescales of impulsive phases are approximately on the order of several minutes \citep{2016duration}, roughly consistent with the timescale ($\sim 5\,\mathrm{min}$) from the rising time of reconnection rate ($t=5.7$) to the final development of turbulence ($t=8.2$) in our simulation.

\section{Discussion and conclusion}\label{Conclusion and Discussion}
Using high-resolution MHD simulation data of 3D flare turbulent reconnection, we investigated the observational characteristics of the current sheet, including the thickness, DEMs, and nonthermal broadening of the Ca \uppercase\expandafter{\romannumeral17} line. 
Our main results include:
\begin{enumerate}
  \item Turbulent magnetic reconnection can significantly broaden the apparent width of the current sheet. The range of apparent widths measured using the synthetic XRT image as seen from the edge-on view is much smaller than that of the physical width of high-temperature structures.
  The minimal and maximal values of the physical widths are much smaller and larger than those of the apparent width, respectively. 
  This is mainly caused by the appearance of turbulence within the current sheet. 
  With the development of turbulent reconnection, various structures of different scales form and are redistributed in the current sheet region.
  The KHI also causes the distortion of the current sheet.
  Together, these effects increase the transverse apparent width of the current sheet.
  The complexity of the current sheet structures and superposition effects make accurately evaluating the current sheet width difficult.
  
  \item The current sheet is significantly heated by reconnection, wherein the high-temperature plasmas along the LOS are co-spatial with small-scale reconnection sites within the current sheet. 
  We obtain DEMs of the current sheet that show two peaks, similar to real observations \cite[see][Fig.\,1C]{2018ApJ...866...64C}. 
  In observations, the low-temperature peak is frequently interpreted to be contributed by the foreground and background in the direction of the LOS \citep{2018ApJ...854..122W}.
  The analysis based on the 3D simulation data, however, shows that the low-temperature component is actually from the plasma that is not heated by the reconnection in the current sheet.
  It is the original plasma brought by the distortion of the flux rope (see the third and fourth columns of Fig.\,\ref{fig:PhysWidthSlice}). 
  Thus, caution should be paid when interpreting low-temperature components of the current sheet DEMs; it might be an imprint of the distortion of the flux rope, which probably has a more important contribution than the foreground or background.
  
  \item Strong turbulence can also cause the nonthermal broadening of spectral lines throughout the current sheet region in addition to in the loop-top region. 
  We synthesized Ca \uppercase\expandafter{\romannumeral17} observations, which exhibit complex spectral line characteristics at different locations with varying velocities and temperatures along the LOS.
  The results for the loop-top region are also basically consistent with those derived by \citet{2023FrASS..1096133S}.
  Compared with the simulation by \citet{2023FrASS..1096133S}, our simulation implements the self-consistent formation of turbulence in the current sheet region with the aid of higher resolution, showing current sheet features highly analogous with observations \citep{2023ApJ...954L..36W}.
  Moreover, we also find that the nonthermal velocities are almost independent of the line intensities and the heights under a fully developed turbulent reconnection state.
\end{enumerate}

There are some limitations in our work. First, we only investigated the properties of the current sheet as viewed edge-on, which provides better comparisons with the observations \citep{2003JGRA..108.1440W,2018ApJ...854..122W} and the 2D standard flare model \citep{2015SSRv..194..237L}, but these properties may vary when viewed from other perspectives.
The edge-on view is the most complex due to the superposition effect.
As the LOS rotates away from the edge-on view, the LOS integral depth across the current sheet will reduce rapidly, decreasing the influence of the superposition effect (see Fig.\,\ref{fig:XRT_MVP}).
Second, we only studied the Ca \uppercase\expandafter{\romannumeral17} line, which presents typical characteristics of the spectral lines within the current sheet, but other spectral lines need to be studied in the future.
Third, the nonthermal broadening does not include the effects of nonthermal particles due to limits of the MHD model.
Fourth, the plasma temperature in the observed current sheet, such as during the SOL2017-09-10T X8.2 flare, can reach above $10\,\mathrm{MK}$ \citep{2018ApJ...866...64C}, but the peak temperature in our simulation is below this value.
One possible reason is that our simulation stops before the peak moment of reconnection, as it is limited by the available computational resources and numerical instability.
However, the main results derived here are supposed to be independent of the peak temperature.
Fifth, our simulation produces a standard flare configuration under a CSHKP model without line-tying at the $z$ boundaries, which can explain the flares with relatively long extended polarity inversion lines. 
For flares with more complex magnetic configurations, such as circular flares \citep{Masson}, the boundary conditions change and might affect the evolution. Further work is needed to investigate the influences of large-scale complex structures on the fine processes of magnetic reconnection in the current sheets.
Finally, limited by the numerical resistivity, the Lundquist number in our simulation is still far smaller than that in the real corona, which can cause an overestimation of the Ohmic heating. To alleviate the heating problem, more sophisticated simulations, for example with higher resolutions, should be considered in the future.

Despite these limitations, our results provide a comprehensive physical origin for the observational characteristics of the current sheet.
It should be emphasized that the present remote-sensing imaging data, as well as spectroscopic devices such as IRIS and EIS \citep{2007SoPh..243...19C,2014IRIS}, have very limited spatiotemporal resolutions that are insufficient to reveal the complex fine structures of the 3D current sheet or to resolve the predicted shear flows along the current sheet that reach $100$ -- $200\,\mathrm{km\,s^{-1}}$ at $t=7.7$ \citep[see][Fig.\,5]{2023ApJ...954L..36W}.
The estimated shear layer width is only $500$ -- $2500\,\mathrm{km,}$ and the shear-flow lifetime is $\sim 200\,\mathrm{s}$.
The fine structures generated by the turbulence are even more difficult to observe.
Therefore, to fully understand the essential physical processes within the flare current sheet, higher spatiotemporal resolution \citep[e.g., with MUSE;][]{DePontieu2022,QuinteroNoda2022} is essential in the future.
%-----------------------------------------------------------------

\begin{acknowledgements}
We thank the referee who provided valuable comments and suggestions that improved and clarified the manuscript. The data analysis in this paper was performed in the cluster system of the High Performance Computing Center (HPCC) of Nanjing University.
This research is supported by the Strategic Priority Research Program of the Chinese Academy of Sciences (Grant No. XDB0560000), the Natural Science Foundation of China (Grant No.
12473057 and 12127901), and the National Key R\&D Program of China (Grant No. 2021YFA1600504).
\end{acknowledgements}

%-------------------------------------------------------------------
\bibliographystyle{aa} % style aa.bst
\bibliography{manuscript} % your references Yourfile.bib

\end{document}